\documentclass[11pt]{article}
\usepackage{blois,epsfig}
\usepackage{amsmath,graphicx,amssymb}

\def\Journal#1#2#3#4{{#1} {\bf #2}, #3 (#4)}

\def\vp{{\bf p}}

\def\be{\begin{equation}}
\def\ee{\end{equation}}
\def\bea{\begin{eqnarray}}
\def\eea{\end{eqnarray}}

\def\H{{\cal H}}

\def\vp{{\varphi}}

\def\cs2{c_{\rm{s}}^2}
\def\U0{{\bar U_0}}
\def\H{{\cal H}}
\newcommand{\fnl}{f_{\text{NL}}}
\newcommand\eq[1]{Eq.~(\ref{#1})}

\begin{document}
\vspace*{4cm}
\title{THE QUANTUM ORIGIN OF COSMIC STRUCTURE}

\author{ K.~A.~MALIK }

\address{
Astronomy Unit,
Queen Mary University of London,
United Kingdom}

\maketitle\abstracts{In this concise, albeit subjective review of
  structure formation, I shall introduce the cosmological standard
  model and its theoretical and observational underpinnings. I will
  focus on recent results and current issues in theoretical cosmology,
  in particular in cosmological perturbation theory and its
  applications. }

%%%%%%%%%%%%%%%%%%%%%%%%%%%%%%%%%%%%%%%%%
\section{Introduction}
\label{sect_intro}
%%%%%%%%%%%%%%%%%%%%%%%%%%%%%%%%%%%%%%%%%

%Recent years have seen a remarkable transformation of cosmology, from
%an rather esoteric branch of theoretical physics and applied
%mathematics (at the borderline/niche between) to WHAT

Recent years have seen a remarkable transformation of cosmology, from
a rather esoteric subject at the borderline of theoretical physics
and applied mathematics to one of the most vibrant and popular areas
of modern Astronomy. This development has been brought about by the
spectacular advance in the subject, such as the development of
cosmological perturbation theory, but more importantly, by the
availability of new observational data sets of unprecedented quantity
{and} quality.

Two data sets are of particular importance in this context, Cosmic
Microwave Background (CMB) experiments Large Scale Structure (LSS)
surveys. The CMB experiments\cite{cmb_exp} , balloon-borne, using dedicated
satellites, or ground-based, have revolutionised our understanding of
the universe, giving us access to information from the very early
universe in form of tiny temperature anisotropies, at the level of $1$
part in $100,000$, imprinted on the CMB.
The exceptionally successful {\sc{Wmap}} satellite mission\cite{wmap}
which started in June 2001 and ended just recently (see contribution
by Hinshaw in this volume), measured the spectrum of these
anisotropies with unprecedented precision on large and medium angular
scales.
The {\sc{Planck}} mission\cite{planck}, launched in May 2009, is now
taking data, and will not only improve temperature anisotropy
measurements and extend them to smaller scales, but will also measure
the polarisation of the CMB (see contribution by de Benardis in this
volume).
LSS surveys improved our understanding of the late
universe. From measuring the positions and redshift distances from
just hundreds or thousands of galaxies, the latest surveys, such as
2df Galaxy Redshift Survey\cite{2df} (now completed), the 6df Galaxy
Survey\cite{6df} and the SDSS\cite{sdss} (both still taking data),
have mapped hundreds of thousands of galaxies, and other objects, and
eventually will have mapped on the order of millions of galaxies.

In the following sections I shall give a brief overview of our current
understanding of structure formation, in particular how vacuum
fluctuations in the fields present in the very universe give rise to
CMB anisotropies and LSS in the late universe. I do apologise in
advance for a rather incomplete and subjective referencing due to the
limited space available.

%%%%%%%%%%%%%%%%%%%%%%%%%%%%%%%%%%%%%%%%%
\section{The evolution of the Universe}
%%%%%%%%%%%%%%%%%%%%%%%%%%%%%%%%%%%%%%%%%

Let us now turn to the evolution of the universe, that the observational
data from CMB experiments and LSS surveys mentioned in the previous
section and our understanding of fundamental physics suggest. In the
following I shall describe the \emph{Cosmological standard model}
reflecting the general consensus in the cosmology community under the
tacit understanding that nothing ``too weird'' is included.

Faced with the observational data, we might first ask what underlying
theory or theories govern the evolution of the Universe, giving rise
to the data. Fortunately we do not have to invoke utterly new physics
to answer this question.
Most of the progress in recent years is built on two familiar theories
from theoretical physics, each governing their particular range of
scales, namely on small scales \emph{Quantum Field Theory}, necessary
to set initial conditions for the universe, and on large scales
\emph{Einstein's General Relativity}, necessary to calculate its
evolution.  This means we can already work with two separate, well
understood theories, instead of having to wait for a more fundamental
final theory.
Although it would be nice to have an underlying more fundamental
theory, we already have the tools to calculate how quantum
fluctuations evolve into large scale structure, thereby testing models
of the early universe which at least reflect some aspects of any more
fundamental theory.

Figure 1 summarises the evolution of the universe in the cosmological
standard model (figure from Faucher-Gigu\`ere et al.{\cite{cosweb}}).
\begin{figure}
\label{fig1}
%\scalebox{.5}
\begin{center}
{ \includegraphics[width=88mm]{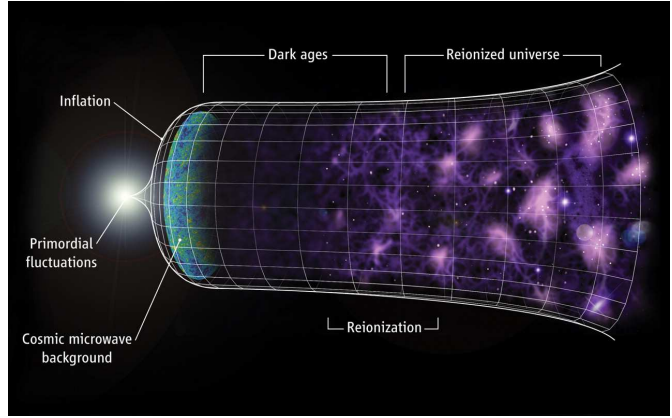} } 
\caption{The evolution of the universe in what has become the
  {\it{cosmological standard model}}.}
\end{center}
\end{figure}
The universe begins with a period accelerated expansion, called
inflation. During this era small quantum fluctuations in the fields
present are stretched by the expansion of spacetime to
super-horizon scales (larger than the particle horizon), where they
are frozen in. These small fluctuations therefore remain constant
until they later reenter the horizon and act as the ``seeds'' for the
anisotropies in the CMB and source the formation of the LSS, where
the small fluctuation are amplified through gravitational instability.
%until they eventually collapse into forming the observable LSS.
%
Roughly $400,000$ years after the beginning of the universe the CMB is
formed, followed by an epoch known as the ``dark ages'' during which
no objects emitting visible light are present. The first stars are
forming ca.~400 Myears afterwards, starting an epoch of reionisation
ending roughly 1 Gyear after the beginning. The first galaxies begin
to assemble during the dark ages, ever larger structures such as
clusters and super-cluster of galaxies form thereafter, and the
universe has begun to undergo another period of accelerated expansion
a couple of Gyears ago.
The main constituents of the universe as given by the seven year WMAP
data (see Komatsu et al.\cite{Komatsu2010}) at the present day are
roughly 73\% dark energy, 23\% dark matter, and 4.5\% baryonic matter.

Let me stress again, unlike the standard model in particle physics,
the cosmological standard model is by no means accepted by the whole
community working in the wider field of cosmology. This reflects on
the one hand the heterogeneity of the field and the community, and on
the other the still weaker experimental and observational
``underpinnings'' of cosmology compared to particle physics, despite
the huge progress in recent years.
%
%Add: might change in future years?
%
The cosmological standard model is however the model most
practitioners might agree upon, albeit some with a slight hesitation.

To understand the final ``details'' -- what drives inflation, what
drives the late time acceleration of the universe -- we might need to
resort to more arcane areas of theoretical physics. I will return to
these questions later on.

%%%%%%%%%%%%%%%%%%%%%%%%%%%%%%%%%%%%%%%%%%%%%%%%%%%%%
\section{Generating primordial density perturbations}
\label{sect_generate}
%%%%%%%%%%%%%%%%%%%%%%%%%%%%%%%%%%%%%%%%%%%%%%%%%%%%%

In this section I shall briefly review how the primordial density
perturbations that source the formation of the CMB anisotropies and
the LSS are generated. As pointed out above, the dynamics of the
universe on large and intermediate scales is controlled by Einstein
equations
\be
\label{Einstein}
G_{\mu\nu}=8 \pi\ G \ T_{\mu\nu}\,,
\ee
where $G_{\mu\nu}$ is the Einstein tensor describing the geometry of
spacetime, $T_{\mu\nu}$ is the energy-momentum tensor encoding the
matter content of the universe, and $G$ is Newton's constant.
However, the initial conditions are set by the theory governing the
smallest scales, Quantum field theory. Of particular importance here
will be the dynamics of the field driving inflation, the \emph{inflaton}.

%%%%%%%%%%%%%%%%%%%%%%%%%%%%%%%%%%%%%%%%%
\subsection{Inflation}
%%%%%%%%%%%%%%%%%%%%%%%%%%%%%%%%%%%%%%%%%

Cosmological inflation was proposed in the early eighties by
Starobinsky\cite{Starobinsky:1980te} and Guth\cite{Guth:1980zm} to
alleviate several problems of the Hot Big Bang model (such as the
flatness, horizon, monopole problems, see e.g.~Liddle and
Lyth\cite{LLBook} for an introduction to modern cosmology).  Inflation
is a period of accelerated expansion of spacetime in the very early
universe. This is ``easily'' achieved by introducing a scalar field
$\vp$ with potential $U$, which gives rise to the pressure
%
%\be
$P=\frac{1}{2a^2}\vp'^2-U(\vp)\,,$
%\ee
% 
where $a$ is the scale factor and a prime denotes differentiation with
respect to conformal time $\eta$.

We can see that the pressure $P$ is negative during a period
of potential domination, which is usually associated with the field
``slowly rolling'' down its potential, therefore having small or negligible
kinetic energy (this is known as the slow-roll approximation).

The dynamics of the background\footnote{I am postponing the details of
  splitting quantities into background and perturbations to Section
  \ref{sect_pert}.} is governed by the Klein-Gordon equation
\be
\label{KG0}
\vp''+2\H \vp'+a^2 U_{,\vp}=0\,,
\ee
for the scalar field, and by the Friedmann equation 
\be
\label{fried}
\H^2=\frac{8\pi G}{3}\left(\frac{1}{2}{\vp'}^2+a^2 U\right)\,,
\ee
for the scalar factor $a$, where $\H=a'/a$. The above system
constitutes a simple damped oscillator. Imposing the slow-roll
approximation then gives a nearly constant Hubble parameter $H=\H/a$.

We now turn to the generation of vacuum fluctuations in the scalar
field. The evolution of these fluctuations is governed by the perturbed
Klein-Gordon equation
\be
\label{KG1}
{\delta\vp_1}''+2\H{\delta\vp_1}'+k^2{\delta\vp_1}
+a^2U_{,\vp\vp}{\delta\vp_1}=0
\ee
where we assumed slow-roll and are working in Fourier space, with $k$
being the comoving wavenumber. Once the potential is specified, \eq{KG1}
can be solved in terms of Hankel functions.

The initial conditions for \eq{KG1} are the connection with Quantum Field
Theory and are imposed on small scales and at early times ($|k\eta|\gg
1$) (choosing the positive frequency modes in the initial vacuum
state, see e.g.~Liddle and Lyth\cite{LLBook}), and are given by
\be
\delta\vp_1\sim\frac{e^{-ik\eta}}{a\sqrt{2k}}\,.
\ee
Then defining the power spectrum for the field fluctuations as
${\cal{P}}_{\delta\vp_1}(k)
\equiv\left(\frac{k^3}{2\pi^2}\right)\big|\delta\vp_1\big|^2$, we get
at horizon crossing, i.e.~when $k=\H$, the now classic result for the
fluctuation amplitude,
${\cal{P}}_{\delta\vp_1}(k)={H^2}/{(2\pi)^2}$. This means that the 
amplitude of the field fluctuations at horizon crossing is (nearly)
independent of scale or scale-invariant for $H\sim const$
(Starobinsky\cite{Starobinsky:1980te}; Hawking\cite{Hawking:1982cz};
Guth and Pi\cite{Guth:1982ec}; Mukhanov and
Chibisov\cite{Mukhanov:1981xt}).
%
%\be
%{\delta\vp_1}\big|_{k=\H}=\frac{H}{2\pi}\,.
%\ee
% 

Inflation solves many problems of the hot Big Bang model which it was
designed to do. Therefore the arguably greatest success of inflation
is the generation of a (nearly) scale-invariant or
Harrison-Zeldovich-Peebles power spectrum\cite{Harrison:1969fb} for
the primordial density fluctuations from the vacuum fluctuations in
the scalar field that can then act as seeds for structure formation,
something it wasn't originally intended to do.

%%%%%%%%%%%%%%%%%%%%%%%%%%%%%%%%%%%%%%%%%
\subsection{Cosmological perturbation theory}
\label{sect_pert}
%%%%%%%%%%%%%%%%%%%%%%%%%%%%%%%%%%%%%%%%%

To accurately calculate the primordial perturbation spectrum and to
relate it to the spectrum of temperature fluctuations in the CMB or
the distribution of galaxies, we need General Relativity.

Unfortunately General Relativity is non-linear, and only a few exact
solution relevant for cosmology are known. We therefore have to resort
to an approximation scheme, which predominantly is cosmological
perturbation theory.

Choosing the homogeneous and isotropic Friedmann-Robertson-Walker
metric as our background, we have to split all metric {and} matter
variables into a time-dependent background and time and space
dependent perturbations, e.g.~for the scalar field above
$\vp=\vp+\delta\vp_1+\frac{1}{2}\delta\vp_2+\ldots$ (the subscripts
denote the order of the perturbation). The perturbed quantities are
then substituted into the governing equations (\ref{Einstein}), and
the resulting expressions truncated at the required order. For example
linear perturbation theory is recovered by neglecting terms of second
order or higher.

Although General Relativity is covariant, splitting variables is not:
spurious gauge modes get introduced and therefore we have to construct
gauge invariant variables, as pioneered by Bardeen\cite{Bardeen80}.
For example, a first order ``coordinate'' transformation
$x^\mu\to\widetilde{x^\mu}=x^\mu+\delta x_1^{~\mu}$,
%
%where $\delta x^{~\mu}=[\delta \eta, \delta x^{~~i}]$,
%
induces a change in the metric variable, here the {curvature perturbation},
and the energy density perturbation as
\be
\widetilde\psi_1=\psi_1+\frac{a'}{a}\delta \eta_1\,,\qquad\qquad
\widetilde{\delta\rho_1}=\delta\rho_1+\rho_0'\delta\eta_1\,,
\ee
where $\delta x_1^{~0}=\delta\eta_1$.
We can now solve for $\delta\eta_1$, combine both equations, and
get a gauge-invariant quantity, which no longer contains any gauge
artefacts,
\be
\label{defzeta}
%\psi_1\Big|_{\delta\rho=0}
-\zeta_1=\psi_1+\frac{\H}{\rho'}\delta\rho_1\,,
\ee
the curvature perturbation on uniform density hypersurfaces (Bardeen
et al.\cite{Bardeen83}). For recent reviews on cosmological
perturbation theory, including higher order perturbations, see
e.g.~Malik and Matravers\cite{MM2008} (more mathematical) and Malik
and Wands\cite{MW2008} (more detailed).

%%%%%%%%%%%%%%%%%%%%%%%%%%%%%%%%%%%%%%%%%
\section{Evolution and conserved quantities}
%%%%%%%%%%%%%%%%%%%%%%%%%%%%%%%%%%%%%%%%%

Variables like $\delta\vp_1$ in general evolve and we need to
model their evolution from the end of inflation, or more precisely, when
they exit the horizon, to the time when they reenter the horizon.
However, we can use instead {conserved quantities}, for which one only
needs to calculate the value at ``horizon exit''.
A popular example is $\zeta_1$ introduced in \eq{defzeta}.
Energy conservation is then sufficient, as shown in Wands et
al.\cite{WMLL}, to guarantee that on large scales for adiabatic
perturbations $\zeta_1'=0\,$.
%
%\be
%\zeta_1'=0\,.
%\ee
% 
We can therefore calculate observable quantities in the early
universe, e.g.~at end of inflation after horizon exit, then map them
onto $\zeta$ and be confident that the observables won't change until
they reenter the horizon.\\

\noindent
Hence we arrive at the following simplistic picture of structure
formation:
\vspace{-3mm}
\begin{itemize}
\setlength{\itemsep}{-3.0pt}
\item
vacuum fluctuation in the scalar field, mapped to the curvature
perturbation $\zeta_1$ $\sim$ gravitational potential wells,
\item
dark matter and other fluids ``fall into'' the potential wells,
amplified by gravitational instability,
\item
CMB anisotropies, and anisotropies in the neutral hydrogen and the LSS
are formed.  
\end{itemize}

%%%%%%%%%%%%%%%%%%%%%%%%%%%%%%%%%%%%%%%%%
\section{Observational signatures}
%%%%%%%%%%%%%%%%%%%%%%%%%%%%%%%%%%%%%%%%%

In order to test our models of the early universe, we have to compare
their theoretical predictions with the observational data. In the
following I shall briefly describe how this is done and which
observable quantities are used.

%%%%%%%%%%%%%%%%%%%%%%%%%%%%%%%%%%%%%%%%%
\subsection{Calculating observational consequences}
%%%%%%%%%%%%%%%%%%%%%%%%%%%%%%%%%%%%%%%%%

As described in Section \ref{sect_generate}, the starting point is the
calculation of the two-point correlator or power spectrum
$\langle\delta\vp_1 \delta\vp_1\rangle$ of the field fluctuations,
which can then be translated into the spectrum of a conserved quantity
that later on source the CMB anisotropies, e.g.~the curvature perturbation
on uniform density hypersurfaces, $\langle\zeta_1 \zeta_1\rangle$.
This input power spectrum has then to be evolved using the Einstein
equations, usually using Boltzmann solvers such as
{\sc{Cmbfast}} or {\sc{Camb}}\cite{cmbfast}, and we get the
theoretical predictions for the CMB anisotropies, which can then be
compared with observational data. The whole process is highly
non-trivial, and beyond the scope of this article.

In comparing the theory with the observations using the formalism
sketched above the theoretical input is in general a particular {model
  of inflation}, given in form of a particular potential $U(\vp)$.
There are too many models to list, and I am following e.g.~Liddle and
Lyth\cite{LLBook} by grouping the model zoo into:
\vspace{-3mm}
\begin{itemize}
\setlength{\itemsep}{-3.0pt}
\item
single field models versus multi-field models,
\item
large field models compared to small field models.
\end{itemize}
It is interesting to note that at present the simplest ``chaotic
inflation'' models, introduced by Linde\cite{Linde1983}, are still in
agreement with the data (following the above categorisation these are
single, large field models), e.g.~$U(\vp)=\frac{1}{2}m^2\vp^2$.\\

Possibly the biggest problem of inflation is that the nature and
identity of the inflaton, the field that drives inflation, is at
present unknown. As indicated above, it is not even clear whether more
than one field is involved, and whether the field driving inflation is
responsible for generating the initial nearly scale invariant power
spectrum.
%\footnote{}

%%%%%%%%%%%%%%%%%%%%%%%%%
%Cosmological parameters\\
%%%%%%%%%%%%%%%%%%%%%%%%%

Let me now very briefly highlight some of the parameter values from
{\sc{Wmap7}} cosmological interpretation paper by Komatsu et
al.\cite{Komatsu2010}.
Taking the  primordial power spectrum as a power law, with amplitude
$\Delta_{\zeta}^2(k_0)$ and spectral index $n_{\rm{s}}$
\be
\Delta_{\zeta}^2(k)
=\Delta_{\zeta}^2(k_0)\left(\frac{k}{k_0}\right)^{n_{\rm{s}}-1}\,,
\ee
we have $\Delta_{\zeta}^2(k_0)=2.43\times 10^{-9}$,
$n_{\rm{s}}=0.969$ (at 68\%C.L.), at pivot scale $k_0=0.002 \ Mpc^{-1}$.
Note that {\sc{Wmap7}} ruled out the exact Harrison-Zeldovich-Peebles
spectrum with spectral index $n_{\rm{s}}=1$ (at more than 3 $\sigma$).
Finally, the scalar to tensor ratio, that is the contribution of gravitational
waves to the power spectrum is $r<0.36$, and the ``running'' or scale
dependence of spectral index is $-0.084< d n_{\rm{s}}/d \ln k<0.010$.
%

%%%%%%%%%%%%%%%%%%%%%%%%%%%%%%%%%%%%%%%%%
\subsection{Higher order observables}
\label{sect_higher_obs}
%%%%%%%%%%%%%%%%%%%%%%%%%%%%%%%%%%%%%%%%%

At linear order in perturbation theory the primordial perturbations
generated during inflation are (very nearly) Gaussian distributed, but
at higher this is no longer the case. Higher order cosmological
perturbation theory has already allowed us to extract new information
from observational data sets and the calculation of new observable
quantities.
Note, that in this article I am mainly concerned with classical
perturbation theory, so the order of the perturbations does not refer
to loops.

As stated above, at linear order the observable of choice is the two-point
correlation function: 
%$\langle \delta\phi\delta\phi \rangle$ 
$\langle \zeta \zeta \rangle$, 
which gives rise to the power spectrum $P(k)\sim A k^{n_{\rm{s}}-1}$,
with amplitude $A$ , spectral index $n_{\rm{s}}$ (and comoving
wavenumber $k$). The power spectrum contains all the information on
the distribution (in the Gaussian case).

At second order in perturbation theory we can calculate the
three-point correlation function, $\langle \zeta\zeta\zeta \rangle$
giving rise to the bispectrum (Gangui et al.\cite{Gangui:1993tt},
Komatsu and Spergel\cite{Komatsu:2001rj}, Maldacena\cite{Maldacena02},
and for a recent review see Bartolo et
al.~\cite{Bartolo:2004if}). This is much more complicated, even in the
Gaussian case.  However, for the simplest models the information in
the bispectrum can be characterised by a single number, the
non-linearity parameter $\fnl$, which in this case can be roughly
described as
%
%\be
$\fnl\propto {\zeta_2}/{(\zeta_1)^2}\,,$
where 
$\zeta_1$ and $\zeta_2$ are the curvature perturbation at first and
second order, respectively.  Note, that at present $\fnl$ is treated
as a constant (as is spectral index in many studies), though
eventually -- when sufficient data is available -- one should allow
for scale and configuration dependence.

Having calculated the theoretical predictions, we can then use the
observational data from the CMB, and increasingly also data from LSS
surveys, to constrain the models of the early universe we are
studying.  At linear order most models under discussion these days
pass the observational tests. The non-linearity parameter $\fnl$ is
therefore becoming a very strong model discriminator: the very
simplest single field inflation models , the ``vanilla'' variety,
predict $\fnl\sim 1$. However, there is already a tantalising hint,
albeit only at the $68\%$ level, in the {\sc{Wmap7}} data that $\fnl =
32\pm 21$.  This implies that the ``vanilla'' inflation models {might
  be} ruled out, and multi-field inflation models, e.g.~curvaton
models or generic multi-field inflation models, might be favoured (see
Alabidi et al.\cite{Alabidi} for a recent discussion of three models
giving large $\fnl$, including the curvaton model).

%%%%%%%%%%%%%%%%%%%%%%%%%%%%%%%%%%%%%%%%%%%%%%%%%%%%%%%%%%%%%%%%
\subsection{New phenomena at higher order in perturbation theory}
%%%%%%%%%%%%%%%%%%%%%%%%%%%%%%%%%%%%%%%%%%%%%%%%%%%%%%%%%%%%%%%%

As sketched in Section \ref{sect_higher_obs} above, higher order
perturbation theory is necessary to exploit the data and calculate
higher order observables. But it can also be used to study new
phenomena that only become apparent beyond the standard, linear
perturbation theory. These higher order phenomena can also be used to
study models of the early universe, allowing to probe different
regions of parameter space. I shall very briefly highlight two effects
at second order, the generation of tensor perturbations, and the
generation of vorticity.

Second order tensor perturbations or gravitational waves are sourced
by a term quadratic in first order density (``scalar'')
perturbations. In the absence of first order gravitational waves,
second order ones can dominate the observational signal on some scales
(see the papers by Ananda et al.\cite{Ananda:2006af} and Baumann et
al.\cite{Baumann:2007zm}).

Whereas vorticity at linear order is assumed to be zero in standard
cosmology, as there are no source terms (such as an anisotropic
pressure), at second order vorticity is sourced by the coupling of
density and entropy perturbations, even in the absence of anisotropic
pressure (see the papers by Christopherson et
al.\cite{Christopherson2009}). This effect might be observable, in
particular on small scales, and should allow for e.g.~additional
bounds on the entropy perturbation.

%%%%%%%%%%%%%%%%%%%%%%%%%%%%%%%%%%%%%%%%%%%%%%%%%%%%
\section{Current and future observational data sets}
%%%%%%%%%%%%%%%%%%%%%%%%%%%%%%%%%%%%%%%%%%%%%%%%%%%%

Until recently, cosmologists preferred the ``clean'' data from the
CMB, which is not affected by complicated astrophysics in the late
universe, to calculate higher order observables. However, at the
moment data sets from the later stages of the universe, namely LSS
surveys and future 21cm anisotropy maps, are becoming another focus of
research (see e.g.~the paper by Komatsu et
al.\cite{Komatsu:2009kd}). This is not only because additional data
sets will deepen our understanding of the universe, but they also
probe different epochs in its evolution.

Whereas LSS surveys, such as the ones mentioned in Section
\ref{sect_intro}, probe redshifts out to a depth of $z\sim 1$, maps of
the neutral hydrogen probe intermediate redshifts
and  have the potential to become an additional source of data.
The neutral hydrogen left over from the Big Bang can be mapped using
its 21cm transition. Primordial perturbations sourcing potential
``wells'' and then generating the temperature anisotropies in the CMB,
later on also source anisotropies in the neutral hydrogen. The 21cm
signal is generated after decoupling but before galaxy formation at
redshift $200 \lesssim z \lesssim 30$ (this can be compared to the
formation of the CMB at decoupling, $z\simeq 1100$). The amount of data
in the 21cm anisotropy maps compared to the CMB is many orders of
magnitude higher (see Loeb and Zaldarriaga\cite{Loeb:2003ya} for
details),
and many 21cm experiments are currently either projected or are already
taking data, e.g.~{\sc{Ska}} and {\sc{Lofar}}. It is however not
clear yet, whether issues such as foreground subtraction can be resolved.

%%%%%%%%%%%%%%%%%%%%%%%%%%%%%%%%%%%%%%%%%
\section{Conclusions}
%%%%%%%%%%%%%%%%%%%%%%%%%%%%%%%%%%%%%%%%%

The cosmological standard model works exceedingly well; inflation,
originally introduced to solve problems of the hot Big Bang model,
provides a mechanism to generate a nearly scale-invariant primordial
spectrum of density perturbations through the vacuum fluctuations in
the fields present in the very early universe, e.g.~the inflaton. The
primordial density perturbations can then act as a source for the CMB
anisotropies and the LSS. New observable quantities, in particular at
higher order in cosmological perturbation theory, and new and better
data will allow to constrain the parameter space for the models of the
early universe further.

However, there are also problems. In particular, \emph{what is the
  inflaton?}  Although no candidate for the inflaton is obvious at
present, at the very least the inflationary paradigm is an excellent
parametrisation of whatever more fundamental theory is sourcing
structure formation. Why the universe has the initial conditions
inferred by the data, indeed, whether it makes sense to ask this
question given our knowledge (or ignorance) at the present time, is
not clear.

The {\sc{Lhc}} in Geneva will probe energies of up to 15 TeV,
recreating conditions last encountered at the very beginning of the
universe. Data from the {\sc{Lhc}} will help to understand the
beginning of the universe and its evolution, and also shed some light
on the many unanswered questions that remain in modern cosmology, such
as on the nature of the dark matter. We therefore have reason to hope
that the forthcoming new results from {\sc{Lhc}} and cosmology will
answer some of our questions about the universe, without raising too
many new ones.

%%%%%%%%%%%%%%%%%%%%%%%%%%%%%%%%%%%%%
\section*{Acknowledgements}
%%%%%%%%%%%%%%%%%%%%%%%%%%%%%%%%%%%%%

I would like to thank the organisers for a very enjoyable and
interesting conference.
KAM is supported, in part, by STFC under Grant ST/G002150/1.

%\section*{Appendix}
% We can insert an appendix here and place equations so that they are
%given numbers such as Eq.~\ref{eq:app}.
%\be
%x = y.
%\label{eq:app}
%\ee

\end{document}